\documentclass[floatfix,twocolumn,aps,prx,reprint,superscriptaddress]{revtex4-1}

\usepackage{physics}
\usepackage{amsmath}
\usepackage{amssymb}
\usepackage{float}
\usepackage{graphicx}
\usepackage{color,soul}
\usepackage{hyperref}
\usepackage[caption=false]{subfig}
\usepackage[colorinlistoftodos]{todonotes}
\usepackage{lipsum}

\usepackage{tikz}
\usepackage{circuitikz}
\usepackage{tikz-timing}
\usetikzlibrary{arrows.meta,decorations.pathmorphing,decorations.pathreplacing,positioning,shapes,fadings}

\newcommand{\BT}[1]{\textcolor{blue}{ #1}}
\newcommand{\RT}[1]{\textcolor{red}{ #1}}

\graphicspath{ {main_figures/} {supplement_figures/} }

\makeatletter
\@ifundefined{textcolor}{}
{%
 \definecolor{BLACK}{gray}{0}
 \definecolor{WHITE}{gray}{1}
 \definecolor{RED}{rgb}{1,0,0}
 \definecolor{GREEN}{rgb}{0,1,0}
 \definecolor{BLUE}{rgb}{0,0,1}
 \definecolor{CYAN}{cmyk}{1,0,0,0}
 \definecolor{MAGENTA}{cmyk}{0,1,0,0}
 \definecolor{YELLOW}{cmyk}{0,0,1,0}
}
\usepackage{dcolumn}
\usepackage{bm}
\usepackage{float}
\makeatother

\begin{document}

\title{
Universal fast flux control of a coherent, low-frequency qubit
}

\author{Helin Zhang}
\affiliation{James Franck Institute, University of Chicago, Chicago, Illinois 60637, USA}
\affiliation{Department of Physics, University of Chicago, Chicago, Illinois 60637, USA}

\author{Srivatsan Chakram}
\affiliation{James Franck Institute, University of Chicago, Chicago, Illinois 60637, USA}
\affiliation{Department of Physics, University of Chicago, Chicago, Illinois 60637, USA}

\author{Tanay Roy}
\affiliation{James Franck Institute, University of Chicago, Chicago, Illinois 60637, USA}
\affiliation{Department of Physics, University of Chicago, Chicago, Illinois 60637, USA}

\author{Nathan Earnest}
\email{Current address: IBM T.J. Watson Research Center, Yorktown Heights, NY 10598, USA}
\affiliation{James Franck Institute, University of Chicago, Chicago, Illinois 60637, USA}
\affiliation{Department of Physics, University of Chicago, Chicago, Illinois 60637, USA}

\author{Yao Lu}
\email{Current address: Department of Applied Physics, Yale University, New Haven, Connecticut 06511, USA }
\affiliation{James Franck Institute, University of Chicago, Chicago, Illinois 60637, USA}
\affiliation{Department of Physics, University of Chicago, Chicago, Illinois 60637, USA}

\author{Ziwen Huang}
\affiliation{Department of Physics and Astronomy, Northwestern University, Evanston, Illinois 60208, USA}

\author{Daniel Weiss}
\affiliation{Department of Physics and Astronomy, Northwestern University, Evanston, Illinois 60208, USA}

\author{Jens Koch}
\affiliation{Department of Physics and Astronomy, Northwestern University, Evanston, Illinois 60208, USA}

\author{David I. Schuster}
\email{Corresponding author: David.Schuster@uchicago.edu}
\affiliation{James Franck Institute, University of Chicago, Chicago, Illinois 60637, USA}
\affiliation{Department of Physics, University of Chicago, Chicago, Illinois 60637, USA}
\affiliation{Pritzker School of Molecular Engineering, University of Chicago, Chicago, Illinois 60637, USA}


\begin{abstract}
The \textit{heavy-fluxonium} circuit is a promising building block for superconducting quantum processors due to its long relaxation and dephasing time at the half-flux frustration point. However, the suppressed charge matrix elements and low transition frequency have made it challenging to perform fast single-qubit gates using standard protocols. We report on new protocols for reset, fast coherent control, and readout, that allow high-quality operation of the qubit with a 14 MHz transition frequency, an order of magnitude lower in energy than the ambient thermal energy scale. We utilize higher levels of the fluxonium to initialize the qubit with $97$\% fidelity, corresponding to cooling it to $190~\mathrm{\mu K}$. We realize high-fidelity control using a universal set of single-cycle flux gates, which are comprised of directly synthesizable fast pulses, while plasmon-assisted readout is used for measurements. On a qubit with $T_1, T_{2e}\sim$~300~$\mathrm{\mu s}$, we realize single-qubit gates in $20-60$~ns with an average gate fidelity of $99.8\%$  as characterized by randomized benchmarking.  
\end{abstract}

\maketitle

\section{Introduction}
Superconducting circuits are among the fastest developing candidates for quantum computers due to steady improvements in coherence times, gate fidelities, and  processor size over the past two decades~\cite{devoret2013superconducting, krantz2019quantum}. These developments have ushered the noisy intermediate scale quantum era~\cite{preskill2018quantum}, and demonstrations of quantum advantage over classical computing~\cite{arute2019quantum}.  Modern superconducting quantum processors are typically based on the transmon circuit, which since its inception has seen improvements in coherence by nearly two orders of magnitude driven largely by decreasing environmental noise~\cite{paik2011observation, gambetta2016investigating, dunsworth2017characterization}. While the transmon circuit has seen widespread use in quantum computation, the fluxonium~\cite{manucharyan2009fluxonium}, introduced a few years later, offers many advantageous properties including a rich level structure, natural protection from charge-noise induced relaxation and dephasing, and reduced sensitivity to flux noise compared with earlier flux qubits~\cite{mooij1999josephson,chiorescu2003coherent}. One of the challenges in making it a building block for larger superconducting processors arises from the slow gates using standard microwave control. In this paper, we demonstrate high-fidelity control of a fluxonium circuit using a universal set of single-cycle flux gates on a qubit whose frequency is an order of magnitude lower than the ambient temperature. In the process, we reimagine all aspects of how the circuit should be controlled and operated, and demonstrate coherence times and gate fidelities that match or exceed those of the best transmon circuits, with the potential for further improvements.

The transmon \cite{koch2007} is one of the simplest in the family of superconducting circuits, realizing a weakly anharmonic oscillator with large dipole matrix elements. This circuit trades off increased sensitivity to decay, and a reduced anharmonicity $\alpha$ for decreased sensitivity to charge-noise induced dephasing.
Despite the maximal susceptibility to relaxation, state-of-the-art transmons have depolarization ($T_1$) times around 100 $\mu$s~\cite{dunsworth2017characterization, nersisyan2019manufacturing, wei2019verifying}, corresponding to quality factors $Q$ of a few million. The gate speeds are, however, limited by the small anharmonicity, typically $\sim 5\%$ of the qubit frequency $\omega_q$, resulting in a theoretical upper bound for the gate infidelity of $\sim \omega_q/\left(Q\alpha\right) \sim 10^{-5}$ and  state-of-the-art values of $\lesssim 1-2 \times 10^{-4}$~\cite{arute2019quantum}. This suggests that gate infidelities can approach $1/Q$ by increasing the anharmonicity in comparison to qubit frequency, and performing gate operations within a few Larmor periods.

The flux qubit~\cite{mooij1999josephson,chiorescu2003coherent,chiorescu2004coherent,yan2016flux}, another member of the superconducting circuit family, already has the desired level structure with a relative anharmonicity $ \alpha/\omega_q \gg 1$. The extreme sensitivity to flux noise of these qubits was mitigated by shunting the Josephson junction with a large superinductor, resulting in the development of the fluxonium \cite{manucharyan2009fluxonium,manucharyan2012evidence,pop2014coherent,vool2014non}. Further improvements in energy relaxation times were obtained by the realization of a heavy-fluxonium \cite{lin2018demonstration, nate2018fluxonium}, which additionally reduced the decay matrix elements using a large shunting capacitor. These variants of the fluxonium are reported to have longer coherence times than transmons in 3d architectures~\cite{vlad2019fluxonium}. Even though the heavy-fluxonium has the desired level structure and large coherence times, fast manipulation of the metastable qubit states remains a challenge due to the suppressed charge matrix elements. While Raman transitions can be used for coherent operations~\cite{nate2018fluxonium, vool2018driving}, these protocols are still relatively slow and require high drive powers, while exposing the qubit to the higher loss rates of excited fluxonium levels involved during the gate. The requirement of fast coherent control thus encourages one to explore new schemes for implementing gates.

In this work, we realize a heavy-fluxonium circuit in a 2d architecture with coherence times $T_1,T_{2e} \sim 300~\mu$s exceeding those of standard transmons. The frequency of the qubit transition is only 14~MHz, an order of magnitude lower than the temperature of the surrounding bath. Therefore, to initialize the qubit we develop and realize a reset protocol that utilizes the readout resonator and higher circuit levels to initialize the qubit with $97\%$ fidelity, effectively cooling the qubit down to $190~\mathrm{\mu}$K. 
Lastly, we use flux pulses to realize high-fidelity single-qubit gates within a single period $2\pi/\omega_q$ of the Larmor oscillation.

\begin{figure}[t]
\centering
\includegraphics[width=\columnwidth]{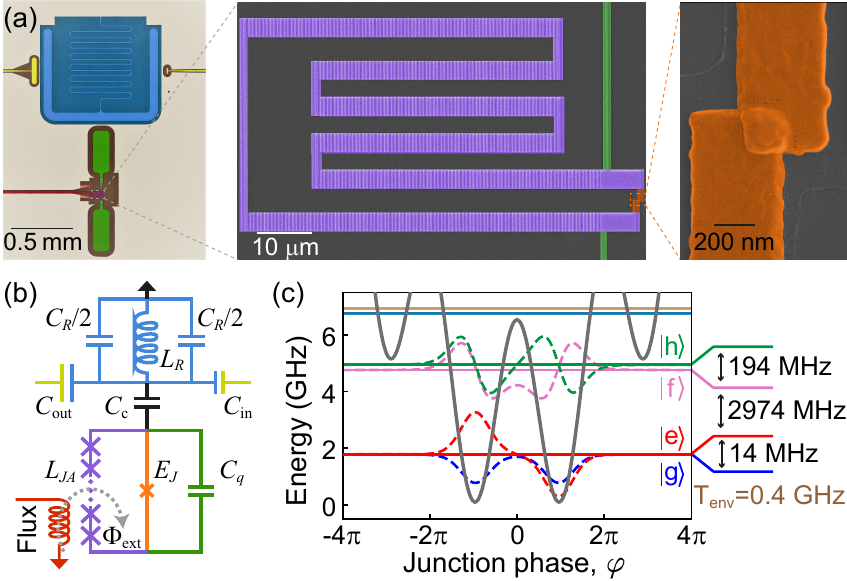}
\caption{Device, circuit and energy level diagram. (a) Left panel: False-colored optical microscope image of the fluxonium coupled to a readout resonator (blue) along with flux (red) and input-output (yellow) lines. Middle panel: scanning electron micrograph of the large junction array (purple), and the small Josephson junction (orange). Right panel: zoom-in view of the small junction. (b) Circuit diagram for the heavy-fluxonium qubit. (c) Energy-level diagram of the heavy-fluxonium at the flux-frustration point ($\Phi_{\rm ext} = \Phi_0/2$). The gray line represents the potential well. The first six energy eigenstates are depicted by the colored lines; dashed lines show the wavefunctions for the first four levels with corresponding color.}
\label{fig:fig1}
\end{figure}

\section{The heavy-fluxonium circuit}

The circuit consists of a small-area Josephson junction (JJ) with inductance $L_J$ shunted by a large inductance ($L_{JA}$), and a large capacitor ($C_q$), as shown in Fig.~\ref{fig:fig1}(a). The shunting inductance is realized by an array of 300 large-area JJs each having a Josephson energy $E_{JA}$ and charging energy $E_{CA}$. We make $E_{JA}/E_{CA}\gg1$ to ensure that the charge dispersion for each array junctions is small, and the array can be regarded as a linear inductor. The corresponding effective circuit is shown in Fig.~\ref{fig:fig1}(b), resulting in a Hamiltonian of the form:
\begin{equation}
   H_f = -4E_C\frac{d^2}{d\varphi^2}-E_J \cos\left(\varphi-2\pi \frac{\Phi_{\mathrm{ext}}}{\Phi_0}\right)+\frac{1}{2}E_L\varphi^2,
\end{equation}
where $E_C=e^2/(2C_q)$ is the charging energy, $E_J=\Phi^2_0/(2L_J)$ the Josephson energy of the small junction, and $E_L = \Phi_0^2/(2L_{JA})$ the inductive energy of the JJ array. $\Phi_{\rm ext}$ denotes the flux threading the loop formed by the small junction and the super-inductance, and $\Phi_0$ is the quantum of flux. The corresponding values for the reported device are: $E_C/h = 0.479$ GHz, $E_L/h = 0.132$ GHz, and $E_J/h = 3.395$ GHz where $h$ is Planck's constant. 
The level structure of fluxonium at the flux-frustration point ($\Phi_{\rm ext} = \Phi_0/2$) is shown in Fig.~\ref{fig:fig1}(c). There are two types of transitions of interest, the intra-well plasmons ($|g\rangle \leftrightarrow |h\rangle$ and $|e\rangle \leftrightarrow |f\rangle$) and inter-well fluxons ($|g\rangle \leftrightarrow |e\rangle$ and $|f\rangle \leftrightarrow |h\rangle$). The single-photon transitions $|g\rangle \leftrightarrow |f\rangle$ and $|e\rangle \leftrightarrow |h\rangle$ are forbidden at the flux-frustration point due to the parity selection rule. The qubit is comprised of the lowest two energy levels $|g\rangle, |e\rangle$, with the qubit transition being fluxon like, 
with a frequency of 14 MHz.

\begin{figure}
\centering
\includegraphics[width=\columnwidth]{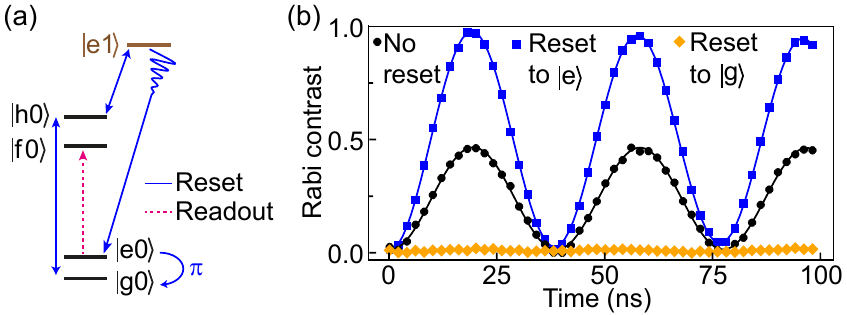}
\caption{
Reset and readout schemes and measurements. (a) Level diagram for the reset and readout protocols. Reset is performed by simultaneously driving both $|g0\rangle \rightarrow |h0\rangle$ and $|h0\rangle \rightarrow |e1\rangle$ transitions (blue double-headed arrows). The spontaneous photon decay $|e1\rangle\rightarrow|e0\rangle$ provides a directional transition (blue single-headed arrow) and completes the reset. An $|e0\rangle \rightarrow |f0\rangle~\pi$ pulse is applied before the readout to boost the output signal. (b) Rabi oscillations between $|e\rangle$ and $|f\rangle$ for different initial state preparations. Blue squares: the initial state is prepared in $\ket{e}$ before the $|e\rangle \leftrightarrow |f\rangle$ Rabi. Black circles: the initial state is the thermal equilibrium state. Orange diamonds: the initial state is prepared in $\ket{g}$.
\label{fig:fig2}}
\end{figure}

\section{Qubit initialization and readout}

Due to its low transition frequency, the qubit starts in a nearly evenly-mixed state in thermal equilibrium. We first initialize the qubit in a pure state ($\ket{g}$ or $\ket{e}$) using the reset protocol shown in Fig.~\ref{fig:fig2}(a). In this protocol, we simultaneously drive both the $|g0\rangle \rightarrow |h0\rangle$ and $|h0\rangle \rightarrow |e1\rangle$ transitions for $15~\mu$s. The high resonator frequency (5.7~GHz) in comparison to the physical temperature, and the low resonator quality factor $Q=600$ result in the rapid loss of a photon from  $|e1\rangle$, effectively removing the entropy from the qubit. In conjunction with the large matrix element between $|h0\rangle$ and $|e1\rangle$, this steers the system into a steady state with over 95\% of the population settling in $|e0\rangle$ in $5~\mu$s (see Appendix~\ref{appendixE}). We subsequently perform an additional $\pi$ pulse on the $\ket{g}-\ket{e}$ transition to initialize the system in the ground state ($|g0\rangle$). The reset is characterized by performing a Rabi rotation between the $|e\rangle \leftrightarrow |f\rangle$ levels, as shown in Fig.~\ref{fig:fig2}(b). The Rabi contrast is doubled following reset, consistent with ~50\% of the population being in $\ket{e}$ in thermal equilibrium. If we prepare the system in $\ket{g}$, the $\ket{e}\leftrightarrow \ket{f}$ Rabi contrast indicates a $3\pm 2\%$ error in state preparation, depending on the $\ket{f}$ state thermal population. Since the $\ket{f}$ frequency is similar to the typical transmon frequencies, its thermal population is in line with that of most transmons. The effective qubit temperature following reset is $\sim 190~\mu$K, lower than the ambient temperature by a factor of 100.

Readout of the fluxonium levels is performed using circuit QED~\cite{wallraff2004strong} by capacitively coupling the fluxonium circuit to a readout resonator~\cite{zhu2013circuit}. Since the qubit states are far away in frequency from the readout resonator, the dispersive shift $\chi$ of the resonator due to a change in the occupation of computational states is small (60~kHz). While the large detuning reduces the qubit heating through the resonator, it makes direct dispersive readout challenging. We circumvent this issue by utilizing the larger dispersive interactions $\chi_f$, $\chi_h$ of the excited levels $|f\rangle$, $|h\rangle$, which are closer in frequency to the readout resonator. 
In order to improve readout fidelity we thus perform a $\pi$ pulse on the $\ket{e}-\ket{f}$ transition in 80~ns, before standard dispersive readout. Since the population in $|e\rangle$ is transferred to $|f\rangle$, the readout signal becomes proportional to ($\chi_f-\chi_g$), which is  5 times larger than ($\chi_e-\chi_g$). This plasmon-assisted readout scheme results in 50\% single-shot readout fidelity, which can be further improved with a parametric amplifier, and by optimizing the resonator $\kappa$ and the dispersive shifts (see Appendix~\ref{appendixF}).

\begin{figure}[b]
\centering
\includegraphics[width=\columnwidth]{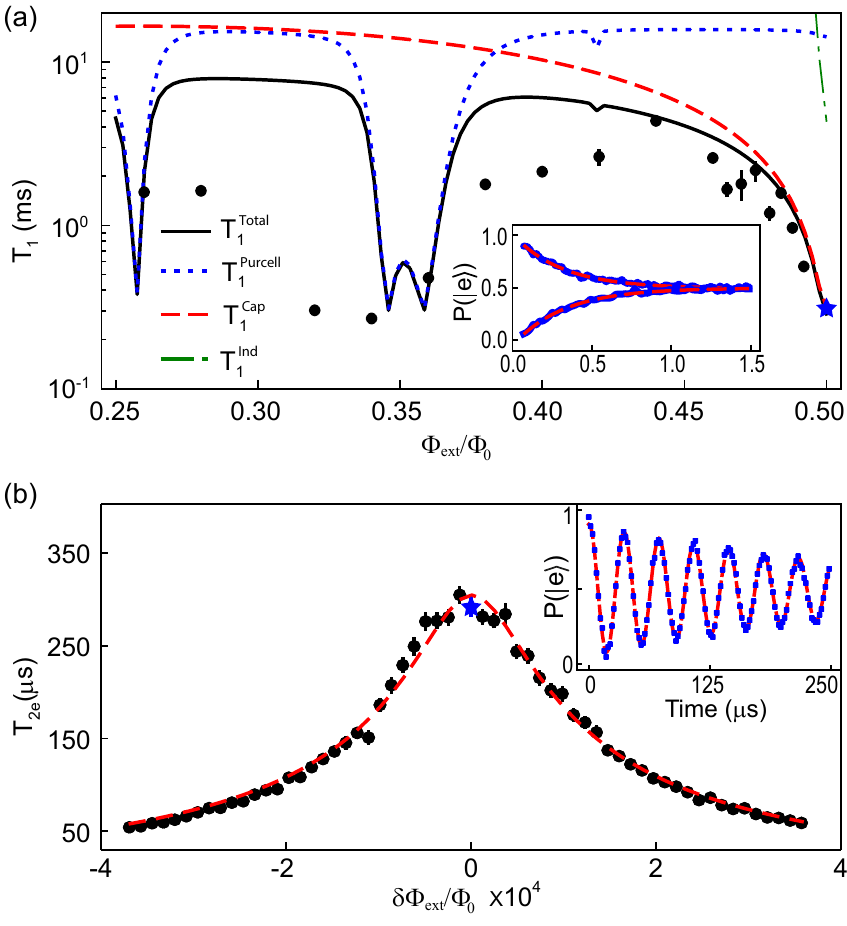}
\caption{
Qubit coherence as a function of flux. (a) Energy relaxation time ($T_1$) as a function of flux, along with the theoretical limits set by dielectric ($T_{1}^{\mathrm{Cap}}$), inductive  ($T_{1}^{\mathrm{Ind}}$), Purcell ($T_{1}^{\mathrm{Purcell}}$), and the combined loss ($T_{1}^{\mathrm{Total}}$). The inset shows the decay of $P(\ket{e})$ to $0.495$ after preparing the qubit in $\ket{g}$,$\ket{e}$ at the flux-frustration point (\BT{$\star$}). (b) Echo decay time $T_{2e}$ as a function of flux near the flux-frustration point. The inset shows an echo measurement at the flux-frustration point (\BT{$\star$}).}
\label{fig:fig3}
\end{figure}

\section{Characterizing device coherence}

Having developed protocols for initialization and readout, we characterize the coherence properties of the qubit. The inset of Fig.~\ref{fig:fig3}(a) shows a $T_{1} =  315 \pm 10~\mu$s  measured at the flux-frustration point following initialization of the qubit in either the $\ket{g}$ or $\ket{e}$ state.  The qubit relaxes to a near equal mixture where the excited state population $P(\ket{e})= 0.4955 \pm 0.0015$, with the deviation providing an estimate of the temperature of the surrounding bath, $T=42 \pm 14$~mK. 
At the flux-frustration point, the wavefunctions are delocalized into symmetric and anti-symmetric combinations of the states in each well. As we move away from this degeneracy point, the wavefunctions localize into different wells resulting in a suppression of tunneling and an increase in the relaxation times, see Fig.~\ref{fig:fig3}(a). Here, the qubit relaxation times were measured over a wide range of external flux by driving the $\ket{g}-\ket{h}$ transition for $120~\mu$s to pump the qubit into the $\ket{e}$ state, and monitoring the subsequent decay. While moving away from the flux-frustration point, $T_{1}$ increases to a maximum value of  $4.3 \pm 0.2$ ms, consistent with previous heavy-fluxonium devices~\cite{nate2018fluxonium,vlad2019fluxonium}, before subsequently decreasing. 

To explain the measured relaxation times, we consider several avenues by which the qubit can decay, including Purcell loss, decay via charge and flux coupling to the control lines, $1/f$ flux noise, dielectric loss in the capacitor, and resistive loss in the superinductor. Conservative estimates of the flux noise induced loss are lower than the measured loss by nearly an order of magnitude (see Appendix~\ref{appendixG}). The loss near the flux-frustration point is believed to be largely due to dielectric loss in the capacitor. This can be thought of as Johnson-Nyquist current noise from the resistive part of the shunting capacitor, which couples to the phase matrix element $\langle g|\hat{\varphi}|e\rangle$, and grows rapidly as we approach the flux-frustration point ~\cite{vlad2019fluxonium}. Assuming a fixed loss tangent for the capacitor, this loss rate is inversely proportional to the impedance of the capacitor, and is given by:
\begin{equation}
  \Gamma_{\rm diel} = \frac{\hbar\omega_q^2}{8 E_C Q_{\rm cap}}\mathrm{coth}\left(\frac{\hbar\omega_q}{2k_BT}\right)|\langle g|\hat{\phi}|e\rangle|^2.
\end{equation}
The $T_1$  at the flux-frustration spot sets an upper bound of $1/Q_{\rm cap} = 8\times10^{-6}$ for the loss tangent of the capacitor, which is within a factor of three of the value reported in previous heavy-fluxonium devices~\cite{vlad2019fluxonium}, and results in the dashed red curve in Fig.~\ref{fig:fig3}(a). Since $\omega_q$ is below the ambient temperature near the flux-frustration point, a combination of the temperature-dependent prefactor $\sim{2k_BT}/(\hbar\omega_q)$, and the relation between charge and phase matrix elements in fluxonium, $\langle g|\hat{n}|e\rangle  = \omega/(8 E_c)\langle g|\hat{\phi}|e\rangle$, results in the dielectric-loss scaling as $1/\omega$, which is consistent with the observed trend in the $T_{1}$ near the flux-frustration point. 
The measured $T_{1}$ at the flux-frustration point also sets an upper bound of $5\times10^{-9}$ for the loss tangent of the inductor. The decay from inductive loss, however, increases more rapidly with frequency than dielectric loss ($\propto 1/\omega^{3}$) and is inconsistent with measured data. Our qubit operations are performed between $0.4\Phi_0-0.5 \Phi_0$ where the $T_1$ is mainly limited by dielectric loss. As we move further away from the flux-frustration point ($\sim 0.4\Phi_0$), $T_1$ starts to decrease. This additional loss is believed to be due to a combination of radiative loss to the charge drive line, and Purcell loss from higher fluxonium levels excited by heating from the $\ket{g}$ and $\ket{e}$ states.  The Purcell loss calculated based on the coupled fluxonium-resonator system using a bath temperature of $60$~mK results in the dotted blue curve shown in Fig.~\ref{fig:fig3}(a). The enhanced loss near $\Phi_{\rm ext} = 0.35 \Phi_0$ is suggestive that heating to higher levels may contribute as there are several near resonances of higher fluxonium levels with the readout resonator, which depend sensitively on the circuit parameters (see Appendix~\ref{appendixG}).

The dephasing is characterized using a Ramsey sequence with three echo $\pi$ pulses, and found to be minimized at $\Phi_{\rm ext} = \Phi_0/2$, where the qubit frequency is first-order insensitive to changes in flux. The dephasing rate near the flux-frustration point can be separated into two parts. The first is a frequency-independent term $\Gamma_C$ mainly composed of qubit depolarization, and dephasing from cavity photon shot noise and other flux insensitive white noise sources. The second arises from $1/f$ flux noise that is proportional to the flux slope as $\Gamma_{1/f} = \frac{d\omega}{d\phi}\eta\sqrt{W}$, where $\eta$ is in the flux-noise amplitude and $W$ depends on the number of $\pi$ pulses in an echo experiment ($W = 4\ln{2}-\frac{9}{4}\ln{3}$ for three $\pi$ pulses~\cite{Ithier2005decoherence}). Thus, our spin-echo signal decays as $\mathrm{exp}(-t/T_C)\times \mathrm{exp}(-\Gamma_{1/f}^2 t^2)$. Here $T_C=1/\Gamma_C$ is the $T_{2e}$ value at the flux-frustration point. It is found to be $\sim 300~\mu$s, much higher than the $T_{2e}$ values for state-of-the-art transmons, see inset of Fig.~\ref{fig:fig3}(b). The $T_{2e}$ values around the flux-frustration point, defined as the time for the echo oscillation amplitude to decay to $1/e$ are shown in Fig.~\ref{fig:fig3}(b). This value falls off rapidly as we move away from the flux-frustration point, consistent with the small tunnel coupling between levels. Away from the flux-frustration point, $T_{2e}$ is mainly limited by $1/f$ flux noise. The $T_{2e}$ far from the frustration point is projected to be $\sim 10~\mu$s according to our model, which is consistent with other reported results~\cite{vlad2019fluxonium}.

\begin{figure}[h]
\centering
\includegraphics[width=\columnwidth]{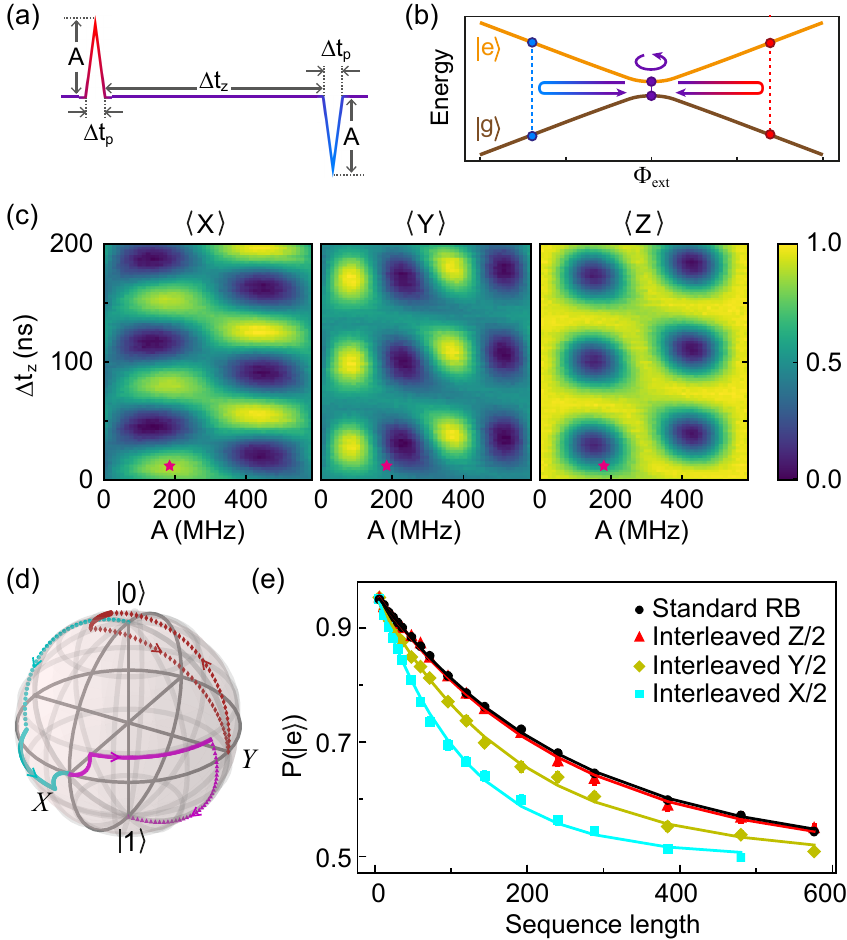}
\caption{
Generic pulse scheme, gate calibration and performance. (a) We use net-zero flux pulses for our native gates ($Y/2$ and $Y$). They are constructed using three sections, a positive triangular pulse with amplitude $A$ and width $\Delta t_p$ on the fast-flux line, an idling period of $\Delta t_z$ and finally another triangular pulse identical to the first one but with a negative amplitude. (b) Energy levels of the computational space as a function of external flux ($\Phi_{\rm ext}$) showing how the fast-flux pulse changes the energies of the instantaneous eigenstates. (c) Expectation values of $\sigma_x, \sigma_y,$ and $\sigma_z$ as a function of pulse parameters $\Delta t_z$ and $A$. These 2D sweeps are used to determine the optimal parameter for the $Y/2$ and $Y$ gates. \RT{$\star$} indicates the parameters for a $Y/2$ gate. (d) Trajectories of three distinct initial states $|0\rangle$ (cyan), $(|0\rangle+|1\rangle)/\sqrt{2}$ (magenta) and $(|0\rangle+i|1\rangle)/\sqrt{2}$ (brown) on the Bloch sphere when a $Y/2$ gate is applied. (e) Comparison of standard RB (black circles) and interleaved RB for $Z/2$ (red triangles), $Y/2$ (gold diamonds) and $X/2$ (cyan squares) gates. The plot is a result of 75 randomized gate sequences averaged over 10000 times. The average gate fidelity is  $\mathcal{F}_{\rm avg} = 0.9980$ and the individual gate fidelities are $\mathcal{F}_{Z/2} = 0.9999$, $\mathcal{F}_{Y/2} = 0.9992$ and $\mathcal{F}_{X/2} = 0.9976$~\cite{Magesan2012IRB}. The uncertainties in all fidelities are smaller than the least significant digit.}
\label{fig:fig4}
\end{figure}

\section{Fast single-cycle flux gates}

In order to maximize the advantage of the large anharmonicity of the heavy-fluxonium, we rethink the standard microwave-drive control of the circuit which is hindered by the suppressed charge matrix elements. We instead perform high-fidelity gates through fast flux pulses, similar to the control scheme used in the original charge qubit~\cite{nakamura1999coherent}. Near the flux-frustration point where the fluxonium is operated, the Hamiltonian within the computational space can be idealized as a spin-1/2 system, $\frac{H}{h} = \frac{A(\Phi_{\rm ext})}{2}\sigma_x + \frac{\Delta}{2}\sigma_z$. Here $\Delta \approx 14$ MHz is the splitting of $\ket{g}$ and $\ket{e}$ at the flux-frustration point, and corresponds to the qubit frequency. The amplitude of the $\sigma_x$ term is proportional to the flux offset $\delta\Phi_{\rm{ext}}
$ from the flux-frustration point, and given by $A = 4\pi\langle g|\hat{\varphi}|e\rangle E_L\delta\Phi_{\rm{ext}}/h$. The coefficient of the $\sigma_x$ term can be much larger than the qubit frequency, with A $\sim300$ MHz when $\delta\Phi_{\rm{ext}} = 0.06\Phi_0$, disallowing any rotating wave approximation.

Fig.~\ref{fig:fig4}(a) shows the protocol for a generic qubit pulse. We first rapidly move the flux-bias point away from the flux-frustration point in one direction and back, thus generating a rotation about the $x$ axis through a large $\sigma_x$ term in our computational basis. There is additionally a relatively small rotation about the $z$ axis corresponding to the time $\Delta t_p$ of the triangular spike. We subsequently idle at the flux-frustration point for a duration $\Delta t_z$, which results in a rotation by $\omega_q\Delta t_z$ about the $z$ axis. Finally, we rapidly move the flux-bias point in the other direction and back, resulting in a -$\sigma_x$ term and another small $z$ rotation. We choose the two spikes to be exactly anti-symmetric, ensuring zero net flux, simultaneously minimizing the  effect  of  microsecond  and  millisecond  pulse  distortions  ubiquitous in flux-bias lines~\cite{DiCarlo2019balanced}, and echoing out low-frequency noise. 
The pulse is also immune to shape distortions since the total $\sigma_x$ and $\sigma_z$ amplitudes depend only on the area of the spike and $\Delta t_z$. By sweeping the amplitude $A$ of the triangular spike and idling length $\Delta t_z$ of the pulse, and measuring the expectation value of the spin along each axis, we obtain the 2d Rabi patterns shown in Fig.~\ref{fig:fig4}(c) that provide a measure of our gate parameters. A vertical line cut of these graphs corresponds to Larmor precession in the lab frame, with an oscillation frequency of $\Delta = 14$ MHz. We thus obtain a  $Z/2$ gate by idling at the flux-frustration point for $\Delta t_z = 1/(4\Delta)$. 
We obtain a $Y/2$ gate at the point indicated by the red star, with the corresponding trajectories on the Bloch sphere for three different cardinal states shown in Fig.~\ref{fig:fig4}(d). $Y/2$ and arbitrary rotations about the $z$ axis are sufficient for universal control. An $X/2$ gate, for instance, is performed through the combination $(-Y/2)\cdot(Z/2)\cdot(Y/2)$.  

We characterize the fidelities of our single-qubit gates through randomized benchmarking (RB) \cite{knill2008randomized,Chow2009RB} and interleaved RB (IRB)~\cite{Magesan2012IRB}.
RB provides a measure of the average fidelity of single-qubit Clifford gates and is performed by applying sequences containing varying number of Clifford gates on the state $\ket{e}$. For a given sequence length, we perform 75 randomized sequences, each containing a recovery gate to the state $\ket{e}$ before the final measurement. IRB allows us to isolate the fidelities of individual computational gates and is performed by interleaving the gate between the random Clifford gates of the RB sequence. The averaged decay curves of $P(\ket{e})$ as a function of the sequence length for standard RB (black circles), and IRB for $Z/2$ (red triangles), $Y/2$ (gold diamonds) and $X/2$ (cyan squares) gates are shown in Fig.~\ref{fig:fig4}(e). The infidelities thus extracted for the $Y/2$, $Z/2$, and $X/2$ gates are $8, 1,~ \rm and ~24\times10^{-4}$, respectively.  The $X/2$ gate infidelity is slightly worse than the combined infidelities from two $Y/2$, and one $Z/2$ gate. The durations for $Y/2$ and $Z/2$ are $\sim 20$~ns, while that for the $X/2$ gate is $\sim 60$~ns, and thus all the computational gates are performed within one qubit Larmor period $2\pi/\omega_q = 70$~ns (see Appendix~\ref{appendixD}), with all the operations occurring in the lab frame. The calculated decoherence limited errors of the $Y/2$, and $X/2$ gates are  ~$6.67\times10^{-5}$ and ~$2\times10^{-4}$, suggesting that the major source of gate error arises from residual calibration errors in the pulse parameters, providing room for improvement even from these state-of-the-art values.

\section{conclusion}

In conclusion, we have realized a  heavy-fluxonium qubit with a 14 MHz transition frequency and coherence times exceeding those of state-of-the-art transmons, while demonstrating protocols for plasmon-assisted reset and readout of the qubit, and a new flux control scheme that performs fast high-fidelity gates.
We have explored a new frequency regime in superconducting qubits and demonstrated the feasibility of a sub-thermal frequency qubit, providing a path for manipulating fluxonium qubits with  computational frequencies in the range of several GHz at temperatures much higher than current dilution-refrigerator temperatures. Our new control scheme has dramatically improved the single-qubit gate speed of fluxonium qubits, making them a viable candidate for large-scale superconducting quantum computation. The gate pulses can be directly synthesized with inexpensive digital to analog converters, and are insensitive to shape distortions. Furthermore, the single-qubit gate scheme used in this work can be generalized to two inductively coupled fluxonium circuits, allowing for two-qubit gate operations without involving the participation of excited levels with more loss. 

\begin{acknowledgments} 
 The authors would like to thank Andrew Oriani for experimental assistance, and Alex Ma, Brendan Saxberg, Alexander Anferov and Jay Lawrence for useful discussions. This work was supported by the Army Research Office under Grant No. W911NF1910016. This work was partially supported by the University of Chicago Materials Research Science and Engineering Center, which is funded by the National Science Foundation under award number DMR-1420709. Devices were fabricated in the Pritzker Nanofabrication Facility at the University of Chicago, which receives support from Soft and Hybrid Nanotechnology Experimental (SHyNE) Resource (NSF ECCS-1542205), a node of the National Science Foundation’s National Nanotechnology Coordinated Infrastructure. 
\end{acknowledgments} 

\appendix
\section{Experimental setup}
\label{appendixA}
\begin{figure}[h]
\centering
\includegraphics[width=\columnwidth]{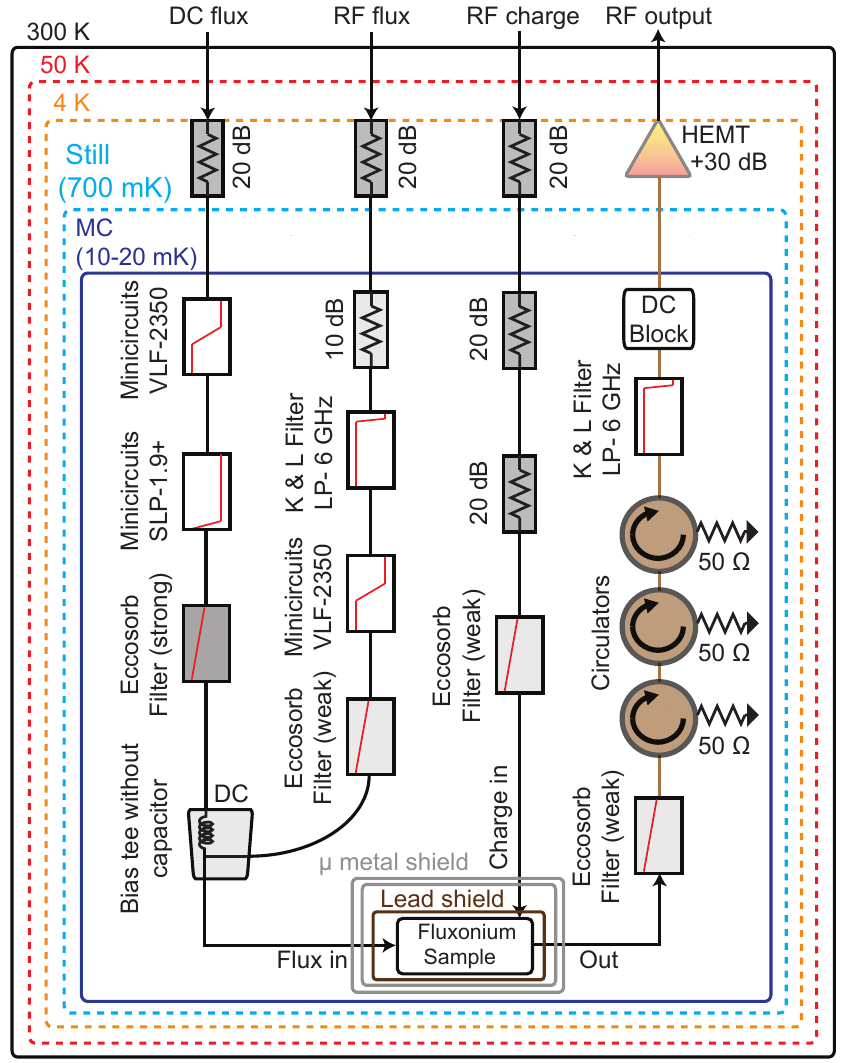}
\caption{Wiring diagram inside the dilution refrigerator. Outside the dilution fridge, there are $\sim16~$ dB of attenuation and a DC block on the RF flux line, and an ultra low pass ($\sim 1~$ Hz) RC filter on the DC flux line. The total attenuation on the RF flux line proved important for both the $T_1$ and $T_2$ of the qubit, likely due to reduction in noise from the Arbitrary waveform generator (Agilent 81180A).}
\label{fig:setup}
\end{figure}
The experiment was performed in a Bluefors LD-250 dilution refrigerator with the wiring configured as shown in Fig.~\ref{fig:setup}. The flux and charge inputs are attenuated at with standard XMA attenuators, except the final 20 dB attenuator on the RF charge line (threaded copper). The DC and RF-flux signals were combined in a modified bias-tee (Mini-Circuits\textsuperscript{\textregistered} ZFBT-4R2GW+), with the capacitor replaced with a short.  The DC and RF-flux lines included commercial low-pass filters (Mini-Circuits\textsuperscript{\textregistered}) as indicated. The RF flux and output lines also had additional low-pass filters with a sharp cutoff (8 GHz) from K\&L microwave. Eccosorb (CR110) IR filters were added on the flux, and output lines, which helped improve the $T_1$ and $T_2$ times, and reduce the qubit and resonator temperatures. The device was heat sunk to the base stage of the refrigerator (stabilized at 15~mK) via an OFHC copper post, while surrounded by an inner lead shield thermalized via a welded copper ring. This was additionally surrounded by two cylindrical $\mu$-metal cans (MuShield), thermally anchored using an inner close fit copper shim sheet, attached to the copper can lid. We ensured that the sample shield was light tight, to reduce thermal photons from the environment.

\section{Device fabrication}
\label{appendixB}
The device (shown in Fig.~1 in the main text) was fabricated on a 430 $\mu$m thick C-plane sapphire substrate. The base layer of the device, which includes the majority of the circuit (excluding the Josephson junctions), consists of 150~nm of niobium deposited via electron-beam evaporation, with features fabricated via optical lithography and reactive ion etch (RIE) at wafer-scale. 600~nm thick layer of AZ MiR 703 was used as the (positive) photoresist, and  the large features were written using a Heidelberg MLA 150 Direct Writer, followed by RIE performed using a PlasmaTherm ICP Fluorine Etch tool. The junction mask was fabricated via electron-beam lithography with a bi-layer resist (MMA-PMMA) comprising of  MMA EL11 and 950PMMA A7. The e-beam lithography was performed on a Raith EBPG5000 Plus E-Beam Writer. All Josephson junctions were made with the Dolan bridge technique. They were subsequently evaporated in Plassys electron beam Evaporator with double angle evaporation ($\pm 19^o$). The wafer was then diced into $7\times7$ mm chips, mounted on a printed circuit board, and subsequently wire-bonded.

\section{Deconstruction of single-qubit gates}
\label{appendixC}
\begin{figure}[t]
\centering
\includegraphics[width=\columnwidth]{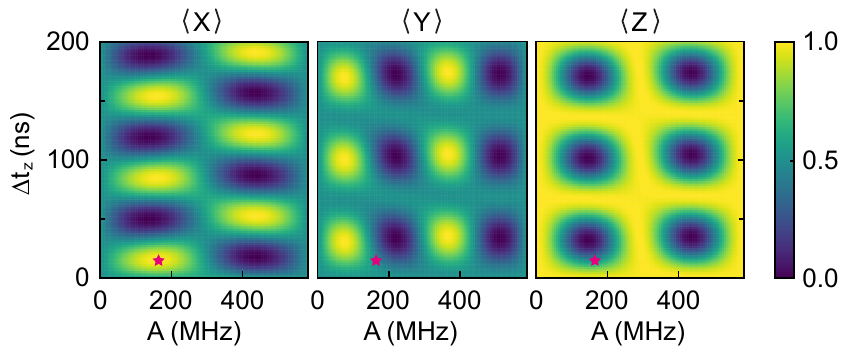}
\caption{Simulated expectation values of $\sigma_x,~\sigma_y$ and $\sigma_z$ as a function of pulse parameters $\Delta t_z$ and $A$ with $\Delta t_p=4.76$~ns. The simulation shows extremely good agreement with the experimental data as shown in Fig.~\ref{fig:fig4}(c).}
\label{fig:sim}
\end{figure}

Modulation of the external flux drive with appropriate amplitude and duration is sufficient to perform arbitrary single-qubit rotations. The native gates available in our system are the arbitrary phase gate $R_z(\theta)$ which rotates the qubit by an arbitrary angle $\theta$ about the $Z$-axis and a combination of $X$- and $Z$-rotation $R_{xz}(\theta)$. $R_z(\theta)$ is realized by waiting for a period of $\Delta t_z = \theta/\omega_q$ (since we are working in the lab frame) whereas $R_{xz}(\theta)$  is implemented by a flux-drive applied for a duration of $\Delta t_p = \lambda \theta/\omega_q$. Here $\lambda$ ($\lambda\le 1$) is the ratio of $Z$-rotation to $X$-rotation rates. These rotation matrices can be expressed as,
\begin{equation}
    R_z(\theta) = e^{-i \sigma_z \theta/2},
\end{equation}
\begin{equation}
\label{eq:Rxz}
    R_{xz}(\theta) = e^{-i(\theta \sigma_x + \lambda |\theta| \sigma_z)/2}.
\end{equation}
The $|\theta|$ in Eq.~\ref{eq:Rxz} arises due to the always-on $Z$-rotation which is unidirectional in the lab frame. A generic zero-flux-pulse can be constructed as,
\begin{equation}
\label{eq:R_generic}
    R(\theta) = R_{xz}(-\theta_x) \cdot R_z(\theta_z) \cdot R_{xz}(\theta_x).
\end{equation}
A $\pi/2$ rotation about the $Y$-axis ($Y/2$), i.e.,
\begin{equation}
    R_y(\pi/2) = \dfrac{1}{\sqrt{2}}
    \begin{pmatrix}
        1 & -1\\
        1 & 1
    \end{pmatrix},
\end{equation}
is obtained using,
\begin{subequations}
\begin{align}
    \theta_x &= \dfrac{1}{\sqrt{1+\lambda^2}} \cos^{-1}\left[ \dfrac{\lambda (1+\lambda)}{-(1-\lambda)} \right], \\
    \theta_z &= 2 \tan^{-1} \left[ \dfrac{\sqrt{1-2\lambda-2\lambda^3-\lambda^4}}{(1+\lambda)\sqrt{1+\lambda^2}} \right]
    \end{align}
\end{subequations}
in Eq.~\ref{eq:R_generic} provided $0\le \lambda \le \sqrt{2}-1$. Similarly, we can construct
\begin{equation}
    R_y(\pi) =
    \begin{pmatrix}
        0 & -1\\
        1 & 0
    \end{pmatrix}
    = -i \sigma_y
\end{equation}
using,
\begin{subequations}
    \begin{align}
        \theta_x &= \dfrac{1}{\sqrt{1+\lambda^2}} \cos^{-1} (\lambda^2), \\
        \theta_z &= \pi - 2\tan^{-1} \left[ \dfrac{\lambda}{\sqrt{1-\lambda^2}} \right],
    \end{align}
\end{subequations}
with $0\le \lambda \le 1$. An arbitrary rotation about $X$-axis can be constructed using,
\begin{equation}
    R_x(\theta) = R_y(\pi/2) \cdot R_z(\theta) \cdot R_y(-\pi/2).
\end{equation}
These gates are sufficient to construct any single-qubit unitary operation. We used the QuTiP \cite{qutip2} python package to simulate the evolution of the computational levels under application of the pulse that was shown in Fig.~\ref{fig:fig3}, and obtained the gate parameters. We swept the drive amplitude $A$ and idling period $\Delta t_z$ in our simulation to match the sweep performed in the experiment, as shown in Fig.~\ref{fig:sim}. $\Delta t_p=4.76$ ns in all the experiments and simulations reported in this paper.

\section{Clifford Gates lengths and fidelities}
\label{appendixD}
\begin{table}[h]
\label{table:gates}
\caption{Clifford gates}
 \begin{tabular}{||c| c| c| c||} 
 \hline
 Gate & Length (ns) & Expm infidelity & Gate Composition \\ [0.5ex] 
 \hline\hline
 $Y/2$ & 21.19 & $8\times10^{-4}$ &  \\ 
 \hline
 $Z/2$ & 17.87 & $1\times10^{-4}$ &  \\
 \hline
 $X/2$ & 60.25 & $24\times10^{-4}$ & $Y/2$, $Z/2$, $-Y/2$ \\
 \hline
 $Y$ & 42.38 &  & $Y/2$, $Y/2$ \\
 \hline
 $Z$ & 35.73 &  & $Z/2$, $Z/2$ \\
 \hline
 $X$ & 78.11 &  & $Y/2$, $Z$, $-Y/2$ \\
 \hline
 
\end{tabular}
\end{table}
A complete Clifford set includes the computational gates $(\exp(\pm i\pi \sigma_j/4), \ j=x,y)$ and the Pauli gates ($\exp(\pm i\pi \sigma_j/2), \ j=I,x,y,z$). In this work, we constructed $Y/2$ and $Z/2$ gates, and used them as building blocks for the other gates in the Clifford Set. The total gate lengths, experimental infidelities (computational gates only), and gate compositions are shown in Table~\ref{table:gates}. The computational gate lengths range from $21-60$~ns, and the longest Pauli gate ($X$) has a length of $78$~ns. Since $2\pi/\omega_q\approx 70$~ns, the computational gates are all within a single cycle of the qubit, and the longest gate is around one cycle as well. The microwave driving gates have lengths longer than $\sim10\times2\pi/\omega_q$, so our gates are $10-30$ times faster.

\section{Fluxonium matrix elements and reset protocol}
\label{appendixE}
We derive the charge drive transition rates by simulating the full qubit-resonator dressed system. The drive power is normalized to $258$~MHz so that the $\ket{g0}\rightarrow\ket{h0}~\pi$ pulse takes $80$~ns, which corresponds to the typical experimental value.  The simulated single photon and 2 photon transition rates (in MHz) are shown in Table~\ref{table:mat_elem_1} and Table~\ref{table:mat_elem_2}. The observed transition rates have additional contributions arising from the frequency dependence of the transmission through the drive line.
\begin{table}[h]
    \caption{One-photon matrix elements}
    \begin{tabular}{|c||c| c| c| c|c|c|} 
    \hline
    & $\ket{g0}$ & $\ket{e0}$ & $\ket{f0}$ & $\ket{h0}$ & $\ket{g1}$ & $\ket{e1}$ \\ 
    \hline\hline
    $\ket{g0}$ & & 0.0738 & & 6.2577 & 257.9425 & \\ 
    \hline
    $\ket{e0}$ & 0.0738 & & 5.8679 & & & 257.9108 \\
    \hline
    $\ket{f0}$ & & 5.8679 & & 1.2475 & 0.0138 &    \\
    \hline
    $\ket{h0}$ & 6.2577 & & 1.2475 & & & 0.1028  \\
    \hline
    $\ket{g1}$ & 257.9425 & & 0.0138 & & & 0.0741  \\
    \hline
    $\ket{e1}$ & & 257.9108 & & 0.1028 & 0.0741 &  \\
    \hline
    \end{tabular}
    \label{table:mat_elem_1}
\end{table}

\begin{table}[h]
    \caption{Two-photon matrix elements}
    \begin{tabular}{|c||c| c| c| c|c|c|} 
    \hline
    & $\ket{g0}$ & $\ket{e0}$ & $\ket{f0}$ & $\ket{h0}$ & $\ket{g1}$ & $\ket{e1}$ \\ 
    \hline\hline
    $\ket{g0}$ & & & 1.9213 & & & 0.9177 \\ 
    \hline
    $\ket{e0}$ & & & & 1.6489 & 0.4207 &  \\
    \hline
    $\ket{f0}$ & 1.9213 & & & & & 0.0644 \\
    \hline
    $\ket{h0}$ & & 1.6489 & & & 0.1258 &  \\
    \hline
    $\ket{g1}$ & & 0.4207 & & 0.1258 & &   \\
    \hline
    $\ket{e1}$ & 0.9177 & & 0.0644 & & & \\
    \hline
    \end{tabular}
    \label{table:mat_elem_2}
\end{table}

\begin{figure}[h]
\centering
\includegraphics[width=\columnwidth]{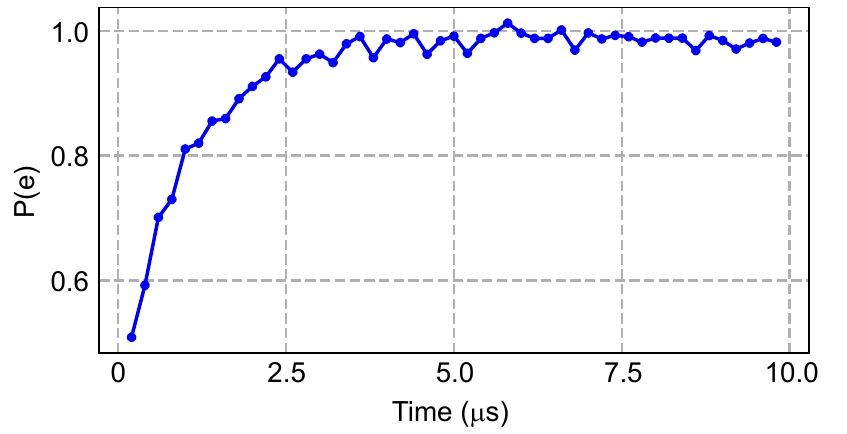}
\caption{The population in the $\ket{e}$ state as a function of the length of the reset pulse. The population is measured after simultaneously driving $\ket{g0}\rightarrow\ket{h0}$ and $\ket{h0}\rightarrow\ket{e1}$ transitions for different lengths of time. Reset of the state is achieved in $\sim 5~\mu$s.}
\label{fig:reset}
\end{figure}

 We utilized the $\ket{g0}\rightarrow\ket{h0}$ and $\ket{h0}\rightarrow\ket{e1}$ transitions for the reset protocol due their large matrix elements. While the $\ket{g0}\rightarrow\ket{e1}$ two-photon process also has a relatively high rate, its use results in deleterious consequences since it lies in the middle of other transitions.
 The excited state population as a function of reset time is shown in Fig.~\ref{fig:reset}. The majority of the population is pumped to state $\ket{e}$ in $5~\mu$s, which is mainly determined by the $\ket{h0}\rightarrow\ket{e1}$ transition rate. We subsequently perform an additional $\pi$ pulse on the $\ket{g}-\ket{e}$ transition to initialize the system in the ground state ($\ket{g0}$).
 
\section{Plasmon assisted readout}
\label{appendixF}
\begin{figure}[h]
\centering
\includegraphics[width=\columnwidth]{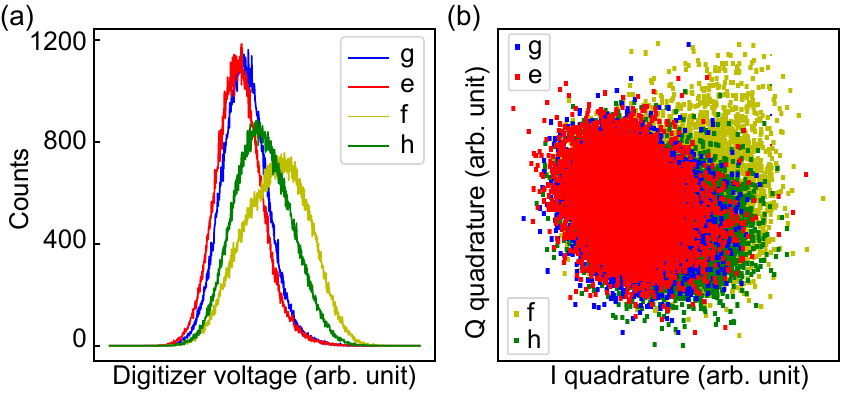}
\caption{Readout histogram and single shot data (a) histogram of the lowest 4 fluxonium states ($\ket{g}$,$\ket{e}$,$\ket{f}$,$\ket{h}$). The $\ket{g}$-$\ket{f}$ readout fidelity is $\sim 50\%$. (b) The distribution of all single shot data from the lowest 4 fluxonium states on the IQ plane.}
\label{fig:hist}
\end{figure}

The resonator frequency shifts in increasing order are $\chi_e, \chi_g, \chi_h, \chi_f$. 
We selected the $\ket{g},\ket{f}$ states for plasmon assisted readout since $\chi_f-\chi_g$ is larger than $\chi_h-\chi_e$. This is reflected in the single-shot readout histogram data for $\ket{g},\ket{e},\ket{f},\ket{h}$ as shown in Fig.~\ref{fig:hist}. 
The histograms are not well separated since the current sample is not optimized for high-fidelity readout.

\section{Modelling fluxonium relaxation}
\label{appendixG}
To explain the measured relaxation times of the fluxonium, we consider  decay via charge and flux coupling to the control lines, $1/f$ flux noise, dielectric loss in the capacitor, resistive loss in the superinductor, and Purcell loss. The decay rates arising from these loss mechanisms are derived using Fermi's golden rule, with the bath described using the Caldeira-Leggett model~\cite{schoelkopf2003qubits,clerk2010introduction}. For a noise source with amplitude $f(t)$ and coupling constant $\alpha$ between the fluxonium qubit states, the interaction Hamiltonian can be written as $H'=\alpha f(t)\sigma_x$ in the qubit subspace. This results in a qubit depolarization rate,
\begin{equation}
\Gamma=\frac{\alpha^2}{\hbar^2}(S_f(+\omega_{01})+S_f(-\omega_{01})).
\label{fermi_golden_rule}
\end{equation}
Here $S_f(\omega) = \int_{-\infty}^{\infty}e^{i\omega\tau}\left<f(\tau)f(0)\right>$ is the noise spectral density associated with the source. We note that at a finite bath temperature corresponding to an inverse temperature $\beta=\frac{1}{k_B T}$, detailed balance relates the positive and negative frequency components of the noise spectral density as
$S_f(-\omega)/S_f(\omega) = e^{-\beta\hbar\omega}$. Depending on the noise source $f$, the coupling constant $\alpha$ is proportional to the charge or phase matrix element of the fluxonium.  Since the only term in the Hamiltonian that does not commute with $\hat{\phi}$ is the charging energy $4E_{c}\hat{n}^2$, and $[\hat{\phi},\hat{n}] = i$,
\begin{eqnarray}
\bra{j}[\hat{\phi},\hat{H}]\ket{k} &=& (\omega_{j}-\omega_{k})\bra{j}\hat{\phi}\ket{k}\nonumber\\
&=& i(8E_{c})\bra{j}\hat{n}\ket{k}. 
\end{eqnarray}
The matrix elements of the fluxonium circuit are thus related by $|\langle g0|\hat{n}|g1\rangle| = (\frac{\omega}{8E_{c}})|\langle g0|\hat{\phi}|g1\rangle|$ for all flux values.

\subsection{Relaxation from flux noise}
Flux noise couples to the phase degree of freedom with an interaction strength that depends on the inductive energy $E_{L}$. Expanding the fluxonium potential to lowest order in flux results in a coupling constant of $\alpha = 2\pi E_{L}\langle g0|\hat{\varphi}|g1\rangle/\Phi_0 $. 
We consider flux noise contributions from current noise in the flux-bias line, as well as $1/f$ flux noise. 
In our experimental setup, the current noise is believed to be mainly due to resistive Johnson-Nyquist noise arising from a 10 dB attenuator with resistance $R = 26~\Omega$ (last resistor in T network) on the fast flux line, corresponding to current noise spectral density of  $S_{I}(\omega)=\frac{2}{R}\frac{\hbar\omega}{\left({1-e^{-\beta\hbar\omega}}\right)}$,  with the expected interpolation between quantum and thermal noise. This is related to flux noise by the mutual inductance $M = \Phi_0/1.6~\mathrm{mA}$ between flux line and the qubit, obtained from the DC flux period. Therefore, $S_f(\omega)+S_f(-\omega)=2\hbar\omega\frac{M^2}{R} \coth{\left(\frac{\beta\hbar\omega}{2}\right)}$, and the decay rate
\begin{equation}
\Gamma_{R} = \pi^{3}\left(\frac{R_Q}{R}\right)\left(\frac{M}{L}\right)^2\langle g0|\hat{\varphi}|g1\rangle|^{2}\omega~\mathrm{coth}\left(\frac{\beta\hbar\omega}{2}\right),
\end{equation}
where $R_{Q} = h/e^{2}$ is the resistance quantum, and $L$ is the fluxonium inductance.

For $1/f$ flux noise, the noise spectral density is of the form $S_\Phi(\omega)=2\pi \eta^{2}/\omega$, with the resulting decay rate,
\begin{equation}
\Gamma_{1/f} = 8\pi^{3}\left(\frac{E_L}{\hbar}\right)^2\left(\frac{\eta}{\Phi_0}\right)^2\frac{|\langle g0|\hat{\varphi}|g1\rangle|^{2}}{\omega}.
\end{equation}
The $1/f$ noise amplitude is fit from $T_{2e}$ data, and corresponds to $\eta=5.21~\mu \Phi_0$. The suppression of the $1/f$ noise induced decay by $E_L^2$, results in a limit of $T_1 = 2.4$ ms for the relaxation time at the flux-frustration point, which grows rapidly ($\propto \omega^{3}$) as we move away from it.

\subsection{Relaxation from radiative loss to the charge line}
In addition to current noise, the fluxonium could also be affected by radiative loss arising from Johnson-Nyquist voltage noise ($S_{V}(\omega)=\frac{2 R\hbar\omega}{{1-e^{-\beta\hbar\omega}}}$) that couples to the qubit via spurious charge coupling, with the resistance $R$ serving as a phenomenological parameter. In this case, the coupling constant is related to the charge matrix element as $\alpha = 2e\langle g0|\hat{n}|g1\rangle$, and $S_f(\omega)+S_f(-\omega)=2R\hbar\omega\coth{\left(\frac{\beta\hbar\omega}{2}\right)}$. The resulting decay rate is,
\begin{equation}
\Gamma_c  =  \frac{\omega}{Q_c}\mathrm{coth}\left(\frac{\beta\hbar\omega}{2}\right)|\langle g0|\hat{n}|g1\rangle|^{2},
\end{equation}
where $Q_c=\frac{R_Q}{16\pi R}$. An upper-bound for the resistance $R$ can be found using the plasmon $T_1$  of  $10~\mu$s, corresponding to a total quality factor of $ 1.86\times10^{5}$, and  $Q_c=7.4\times 10^4$. This results in a fluxon $T_1$ limit in excess of 60 ms at the flux-frustration point. 

\subsection{Relaxation from dielectric loss in the capacitor}
Dielectric loss associated with the capacitor can be thought of as Johnson-Nyquist current noise from the resistive part of the shunting capacitor, which couples to the phase matrix element ($\langle g|\hat{\varphi}|e\rangle$). This loss rate is therefore inversely proportional to the impedance of the capacitor, assuming a fixed loss tangent ($1/Q_{\rm diel}$) for the capacitor. As a result, $S_f(\omega)+S_f(-\omega)=\frac{\hbar\omega^2C}{Q_{\rm diel}}\coth{\left(\frac{\beta\hbar\omega}{2}\right)}$, and 
\begin{equation}
\Gamma_{\rm diel}  =  \frac{\hbar\omega^2}{8E_CQ_{\rm cap}}\mathrm{coth}\left(\frac{\beta\hbar\omega}{2}\right)|\langle g0|\hat{\phi}|g1\rangle|^{2}.
\end{equation}
If the $T_1$ at the frustration point were limited by dielectric loss, a bath temperature of $42$~mK would result in $Q_{\rm cap}=1/(8\times 10^{-6})$. This is close to the expected loss tangent and within a factor of two of that observed in similar fluxonium devices~\cite{vlad2019fluxonium}. This is believed to be the dominant loss channel near the frustration point, also capturing the flux/frequency dependence of the measured loss ($\propto 1/\omega$).
\subsection{Relaxation from resistive loss in the inductor}
For inductive loss, we again assume a frequency independent loss tangent ($L \rightarrow L(1+i/Q_{\rm ind})$), resulting in Johnson-Nyquist current noise that is inversely proportional to the impedance of the superinductor, i.e.,
$S_f(\omega)+S_f(-\omega)=\frac{\hbar}{LQ_{\rm ind}}\coth{\left(\frac{\beta\hbar\omega}{2}\right)}$. The inductive loss is thus,
\begin{equation}
\Gamma_{\rm ind}  =  \frac{E_L}{\hbar Q_L}\mathrm{coth}\left(\frac{\beta\hbar\omega}{2}\right)|\langle g0|\hat{\phi}|g1\rangle|^{2}.
\end{equation}
The superinductor is extremely low loss, with a quality factor of $Q_{\rm ind} = 5\times 10^{9}$ resulting in a limit of $T_1 = 2$ ms at the flux frustration point, growing as $\omega^{3}$ as we move away from the flux-frustration point.

\subsection{Relaxation rate due to the Purcell Effect}

We derive the Purcell relaxation rates of the fluxonium levels, arising from coupling to the resonator by closely following Ref.~\cite{2018Groszkowski}. We model this by assuming that the resonator is coupled to a bath of harmonic oscillators, whose Hamiltonian reads
\begin{align}
H_{\text{bath}}=\sum_{k}\hbar\omega_{k}b_{k}^{\dagger}b_{k},
\end{align}
where $b_{k}$ is the lowering operator for mode $k$. The interaction Hamiltonian between the bath and the resonator is given by 
\begin{align}
H_{\text{int}}=\hbar\sum_{k}\lambda_{k}(ab_{k}^{\dagger}+a^{\dagger}b_{k}),
\end{align}
where $a$ is the lowering operator for the resonator. Finally, the system under consideration is the fluxonium circuit coupled to the resonator, which we write in the dressed basis as
\begin{align}
H_{\text{flux+res}}=\sum_{k}E_{k}^{\text{flux+res}}\ket{\psi_{k}^{\text{flux+res}}}\bra{\psi_{k}^{\text{flux+res}}}.
\end{align}
We treat $H_{\text{int}}$ as a perturbation which can induce transitions among the eigenstates of the Hamiltonian $H=H_{\text{bath}}+H_{\text{flux+res}}$, given by
\begin{align}
\ket{\psi_{i}}=\ket{\psi_{i}^{\text{flux+res}}}\bigotimes_{k}\ket{m_{k}}.
\end{align}
The transition rate under the action of a constant perturbation is given by Fermi's Golden Rule in the form
\begin{align}
\label{FGR}
\gamma_{i\rightarrow f}=\frac{2\pi}{\hbar}\delta(E_{i}-E_{f})|\mel{\psi_{f}}{H_{\text{int}}}{\psi_{i}}|^2,
\end{align}
where $E_{i}$ and $E_{f}$ are the eigenenergies of the states $\ket{\psi_{i}}$ and $\ket{\psi_{f}}$, respectively. These energies are
\begin{align}
E_{i}&=E_{i}^{\text{flux+res}}+\hbar\sum_{k}m_{k}\omega_{k}, \\ \nonumber
E_{f}&=E_{f}^{\text{flux+res}}+\hbar\sum_{k}m_{k}'\omega_{k},
\end{align}
where $\{m_{k}\}$ denotes the initial configuration of the bath and $\{m_{k}'\}$ the final configuration. Inserting the form of $H_{\text{int}}$ into Eq.~\eqref{FGR} and noting that cross-terms vanish leads to
\begin{widetext}
\begin{align}
\gamma_{i,\{m_{k}\}\rightarrow f,{\{m_{k}'\}}}=2\pi\hbar\delta(E_{i}-E_{f})  
\sum_{k} |\lambda_{k}|^2\Big(&\left|\mel{\psi_{f}^{\text{flux+res}}}{a^{\dagger}}{\psi_{i}^{\text{flux+res}}}\right|^2 m_{k}\delta_{m_{k}', m_{k}-1} \\ \nonumber +&\left|\mel{\psi_{f}^{\text{flux+res}}}{a}{\psi_{i}^{\text{flux+res}}}\right|^2 (m_{k}+1)\delta_{m_{k}', m_{k}+1}\Big)\prod_{k'\neq k}\delta_{m_{k'}',m_{k'}},
\end{align}
\end{widetext}
 To find the total transition rate, we must sum over all such initial and final configurations, taking into account the thermal probability of occupying a given initial configuration:
\begin{align}
\Gamma_{i\rightarrow f}=\sum_{\{m_{k}\},\{m_{k}'\}}P(\{m_{k}\})\gamma_{i,\{m_{k}\}\rightarrow f,{\{m_{k}'\}}}, 
\end{align}
where
\begin{align}
P(\{m_{k}\})=\frac{e^{-\sum_{k}\beta m_{k}\hbar\omega_{k}}}{Z},
\end{align}
$Z$ is the partition function of the bath and $\beta=1/k_{B}T$. Performing the sums over all initial and final states yields

\begin{align}
\Gamma_{i\rightarrow f}=2\pi\hbar\sum_{k}|\lambda_{k}|^2\delta(E_{i}^{\text{flux+res}}-E_{f}^{\text{flux+res}}+\hbar\omega_{k}) \\ \nonumber \left|\mel{\psi_{f}^{\text{flux+res}}}{a^{\dagger}}{\psi_{i}^{\text{flux+res}}}\right|^2n_{\text{th}}(\omega_{k}) \\ \nonumber
+2\pi\hbar\sum_{k}|\lambda_{k}|^2\delta(E_{i}^{\text{flux+res}}-E_{f}^{\text{flux+res}}-\hbar\omega_{k}) \\ \nonumber \left|\mel{\psi_{f}^{\text{flux+res}}}{a}{\psi_{i}^{\text{flux+res}}}\right|^2(n_{\text{th}}(\omega_{k})+1),
\end{align}
where 
\begin{align}
n_{\text{th}}(\omega_{j})=\sum_{\{m_{k}\}}P(\{m_{k}\})m_{j}
=\frac{1}{e^{\beta\hbar\omega_{j}}-1}.
\end{align}
We next take the continuum limit and define $\kappa=2\pi\hbar\rho(\omega_{k})|\lambda_{k}|^2$, where $\rho(\omega)$ is the density of states of the bath. Introducing $\omega_{jj'}^{\text{flux+res}}=(E_{j}^{\text{flux+res}}-E_{j'}^{\text{flux+res}})/\hbar$ leads to the expressions
\begin{align}
\Gamma_{i\rightarrow f}^{\uparrow} = \kappa n_{\text{th}}(\omega_{fi}^{\text{flux+res}})\left|\mel{\psi_{f}^{\text{flux+res}}}{a^{\dagger}}{\psi_{i}^{\text{flux+res}}}\right|^2,
\end{align}
for upward transitions $E_{f}^{\text{flux+res}}>E_{i}^{\text{flux+res}}$, and 
\begin{align}
\Gamma_{i\rightarrow f}^{\downarrow} = \kappa (n_{\text{th}}(-\omega_{fi}^{\text{flux+res}})+1)\left|\mel{\psi_{f}^{\text{flux+res}}}{a}{\psi_{i}^{\text{flux+res}}}\right|^2,
\end{align}
for downward transitions $E_{f}^{\text{flux+res}}<E_{i}^{\text{flux+res}}$. The final step is to note that throughout this experiment, the fluxonium qubit is operated in the dispersive regime with respect to the frequency of the resonator. Therefore, we expect that the dressed eigenstates of $H_{\text{flux+res}}$ can be labeled with quantum numbers $\ell$ and $n$, with $\ell$ labeling the fluxonium state and $n$ the resonator state. When performing numerical simulations, this identification is based on which numbers $\ell$ and $n$ produce the maximum overlap of the dressed state $\ket{\psi_{i}^{\text{flux+res}}}=\ket{\overline{\ell,n}}$ with the product state $\ket{\ell,n}$. As in Ref.~\cite{2018Groszkowski}, we are interested mainly in transitions among fluxonium states, where the quantum number $\ell$ changes. We therefore define the total transition rate due to the Purcell effect among fluxonium states as a sum over all possible initial and final states of the resonator, weighting initial states by their probability of being thermally occupied $P_{\text{res}}(n)=(1-\exp(-\beta\hbar\omega_{r}))\exp(-n\beta\hbar\omega_{r})$. This yields
\begin{align}
\Gamma_{\ell\rightarrow\ell'}^{\text{Purcell}, \uparrow} = \sum_{n,n'}&P_{\text{res}}(n)\kappa n_{\text{th}}(\omega_{\ell',n',\ell,n}) \\ \nonumber &\times\left|\mel{\overline{\ell',n'}}{a^{\dagger}}{\overline{\ell,n}}\right|^2,
\end{align}
for upward transitions, where $\omega_{\ell',n',\ell,n}=(E_{\ell',n'}-E_{\ell,n})/\hbar$, and
\begin{align}
\Gamma_{\ell\rightarrow\ell'}^{\text{Purcell}, \downarrow} = \sum_{n,n'}&P_{\text{res}}(n)\kappa (n_{\text{th}}(-\omega_{\ell',n',\ell,n})+1)\\ \nonumber &\times \left|\mel{\overline{\ell',n'}}{a}{\overline{\ell,n}}\right|^2,
\end{align}
for downward transitions. 

The direct Purcell loss ($\ket{e}\rightarrow\ket{g}$) gives a $T_1$ limit $\sim 100$ ms, effectively negligible in our experiments. However, heating to the excited levels of fluxonium due to the finite bath temperature, results in enhanced Purcell loss.  Some of these states ($8^{\rm th}, 9^{\rm th}$ and $10^{\rm th}$ eigenstates) have transition frequencies from the logical manifold that are close to the resonator frequency, resulting in avoided crossings. While their exact location depends sensitively on the circuit parameters, these resonances are likely responsible for the decreased $T_1$ observed near $0.35~\Phi_0$. The total Purcell relaxation rate for a bath temperature of $60$ mK corresponds the dotted blue curve in Fig.~\ref{fig:fig3}(a) of the main text.

\section{Modelling fluxonium dephasing}
\label{appendixH}
On the flux slope, the decay envelope of a Ramsey experiment is best approximated by a gaussian $\mathrm{exp}(-t^2/T_\phi^2)$, where $T_\phi=\Gamma_\phi^{-1}=(\sqrt{2}\eta(\partial_\phi \omega_{01})\sqrt{\ln{\omega_{ir} t}})^{-1}$ to first order. For the spin-echo experiments, low-frequency noise has a reduced weight in the noise spectrum, with $T_\phi=(\sqrt{W}\eta(\partial_\phi \omega_{01}))^{-1}$. We can calculate $W$ for three echo $\pi$ pulses based on ~\cite{Ithier2005decoherence}. At the flux frustration point, the qubit is first order insensitive to $1/f$ flux noise, and the spin-echo data can be explained with an exponential decay from white noise ($T_{2e}=T_C=\Gamma_C^{-1}$). In the regime of our spin-echo flux sweep, both noise sources contribute significantly. The data is therefore fit to a product of a gaussian and an exponential~\cite{2018Groszkowski}, with the $T_{2e}$ defined as  $\mathrm{exp}(-T_{2e}/T_C-T_{2e}^2/T_\phi^2)=1/e$, i.e.,
\begin{equation}
    T_{2e} = \frac{\sqrt{1/T_C^2+4/T_\phi^2}-1/T_C}{2/T_\phi^2}.
\end{equation}

\bibliography{thebibliography}

\end{document}